\documentclass[english,aps,citeautoscript,elide,prb,showpacs,superscriptaddress,twocolumn]{revtex4-1}
\usepackage{mathptmx}
\usepackage[T1]{fontenc}
\usepackage[latin9]{inputenc}
\usepackage{color}
\usepackage{babel}
\usepackage{prettyref}
\usepackage{amsmath}
\usepackage{amssymb}
\usepackage{graphicx}
\usepackage{esint}
\usepackage[unicode=true,
 bookmarks=true,bookmarksnumbered=false,bookmarksopen=true,bookmarksopenlevel=3,
 breaklinks=true,pdfborder={0 0 0},backref=false,colorlinks=true]
 {hyperref}
\hypersetup{pdftitle={Interference of stimulated electronic Raman scattering and linear absorption in coherent control},
 pdfauthor={Julien Rioux},
 pdfsubject={coherent control; current injection; electronic Raman effect; graphene},
 citecolor=blue,linkcolor=blue,pdfpagemode=UseNone,pdfstartview=FitH,urlcolor=blue}

\makeatletter
 
 \@ifundefined{textcolor}{}
 {%
   \definecolor{BLACK}{gray}{0}
   \definecolor{WHITE}{gray}{1}
   \definecolor{RED}{rgb}{1,0,0}
   \definecolor{GREEN}{rgb}{0,1,0}
   \definecolor{BLUE}{rgb}{0,0,1}
   \definecolor{CYAN}{cmyk}{1,0,0,0}
   \definecolor{MAGENTA}{cmyk}{0,1,0,0}
   \definecolor{YELLOW}{cmyk}{0,0,1,0}
 }

\usepackage{afterpage}
\usepackage{hypcap}
\usepackage{svn}
\usepackage{times}
\usepackage{upgreek}

\SVN $Date: 2014-09-22 16:48:47 -0400 (Mon, 22 Sep 2014) $
\SVN $Revision: 713 $

\renewcommand{\vec}[1]{\boldsymbol{\mathbf{#1}}}

\let\citealp\onlinecite

\AtBeginDocument{\renewcommand{\eqref}[1]{\hyperref[#1]{(\ref*{#1})}}}
\newrefformat{eq}{\hyperref[#1]{Eq.~(\ref*{#1})}}
\newrefformat{eqs}{Eqs.~\hyperref[#1]{(\ref*{#1})}}
\newrefformat{equation}{\hyperref[#1]{Equation~(\ref*{#1})}}
\newrefformat{fig}{\hyperref[#1]{Fig.~\ref*{#1}}}
\newrefformat{figure}{\hyperref[#1]{Figure~\ref*{#1}}}
\newrefformat{sec}{\hyperref[#1]{Sec.~\ref*{#1}}}
\newrefformat{subfiga}{\hyperref[#1]{Fig.~\ref*{#1}(a)}}
\newrefformat{subfigb}{\hyperref[#1]{Fig.~\ref*{#1}(b)}}
\newrefformat{subfigc}{\hyperref[#1]{Fig.~\ref*{#1}(c)}}
\newrefformat{subfigd}{\hyperref[#1]{Fig.~\ref*{#1}(d)}}
\newrefformat{subfige}{\hyperref[#1]{Fig.~\ref*{#1}(e)}}
\newrefformat{subfigf}{\hyperref[#1]{Fig.~\ref*{#1}(f)}}
\newrefformat{tab}{\hyperref[#1]{Table~\ref*{#1}}}

\allowdisplaybreaks

\makeatother

\begin{document}

\preprint{Revision \SVNRevision}

\title{Interference of stimulated electronic Raman scattering and linear
absorption in coherent control}

\author{J.~Rioux}

\affiliation{Department of Physics, University of Konstanz, D-78457 Konstanz,
Germany}

\author{J.~E.~Sipe}

\affiliation{Department of Physics and Institute for Optical Sciences, University
of Toronto, 60 St.~George Street, Toronto, Ontario, Canada M5S~1A7}

\author{Guido Burkard}

\affiliation{Department of Physics, University of Konstanz, D-78457 Konstanz,
Germany}

\date{\SVNDate}
\begin{abstract}
We consider quantum interference effects in carrier and photocurrent
excitation in graphene using coherent electromagnetic field components
at frequencies $\omega$ and $2\omega$. The response of the material
at the fundamental frequency $\omega$ is presented, and it is shown
that one-photon absorption at $\omega$ interferes with stimulated
electronic Raman scattering (combined $2\omega$ absorption and $\omega$
emission) to result in a net contribution to the current injection.
This interference occurs with a net energy absorption of $\hbar\omega$
and exists in addition to the previously studied interference occurring
with a net energy absorption of $2\hbar\omega$ under the same irradiation
conditions. Due to the absence of a bandgap and the possibility to
block photon absorption by tuning the Fermi level, graphene is the
perfect material to study this contribution. We calculate the polarization
dependence of this all-optical effect for intrinsic graphene and show
that the combined response of the material at both $\omega$ and $2\omega$
leads to an anisotropic photocurrent injection, whereas the magnitude
of the injection current in doped graphene, when transitions at $\omega$
are Pauli blocked, is isotropic. By considering the contribution to
coherent current control from stimulated electronic Raman scattering,
we find that graphene offers tunable, polarization sensitive applications.
Coherent control due to the interference of stimulated electronic
Raman scattering and linear absorption is relevant not only for graphene
but also for narrow-gap semiconductors, topological insulators, and
metals.
\end{abstract}

\pacs{73.50.Pz, 42.65.Dr, 78.67.Wj, 42.65.-k}

\keywords{coherent control; Dirac electrons; electronic Raman effect; elemental
semiconductors; graphene; nondegenerate optical Kerr effect; nonlinear
optics; optical Kerr effect; optical properties; photoconductivity;
quantum interference phenomena; Raman spectra; stimulated Raman scattering;
two-photon processes}

\maketitle

\global\long\def\bra#1{\left\langle #1\right|}
\global\long\def\ket#1{\left|#1\right\rangle }

\global\long\def\cc{\mathrm{c.c.}}
\global\long\def\d{\mathrm{d}}
\global\long\def\Hamiltonian{H}
\global\long\def\Heaviside{\Theta}
\global\long\def\phase{\varphi}
\global\long\def\disparity{d}

\setcounter{topnumber}{0}
\afterpage{\setcounter{topnumber}{1}}

\section{Introduction}

Coherent control (CC) involves the interference between multiple excitation
pathways to acquire a handle on the final state of a quantum-mechanical
process \citep{Kral2007}. The field originates from progress in manipulating
the transition rates of multi-photon molecular processes \citep{Brumer1986,*Shapiro1988,*Shapiro2003,Warren1993,*Rabitz2000}
and has since grown to encompass condensed-matter systems, including
bulk and nanostructured semiconducting materials \citep{Dupont1995,Atanasov1996,*Hache1997,Sheik-Bahae1999,Manykin2001,Marti2005a,Rodrigo2013},
metal-semiconductor heterostructures \citep{Thunich2012}, and optical
lattices \citep{Zhuang2013}. Applied to semiconductor optics, the
typical CC experiment uses light at a fundamental frequency $\omega$
and its second harmonic $2\omega$. Interference between the transition
amplitudes for two-photon absorption (2PA) at $\omega$ and one-photon
absorption (1PA) at $2\omega$ results in an unbalanced distribution
of carriers in momentum space, providing a net photocurrent \citep{Dupont1995,Atanasov1996}.
The two-color electromagnetic field provides the energy necessary
to create electron-hole pairs in the sample, and the injected charge
carriers acquire their velocities according to the band dispersion
\citep{Sheik-Bahae1999}.

In the typical operation regime for conventional semiconductors such
as GaAs, $\hbar\omega$ would lie within the bandgap; the light frequencies
in such experiments are regularly selected according to $\hbar\omega<E_{g}<2\hbar\omega$,
where $E_{g}$ is the bandgap energy, so that 1PA at $\omega$ is
energetically forbidden, and the lowest energetically-allowed perturbative
order for absorption of the fundamental is the second order. However,
for $\hbar\omega>E_{g}$, the fundamental is absorbed at first order,
and how this affects the CC is especially important to understand
for narrow-gap semiconductors. Gapless materials such as graphene
further present the possibility of studying CC in interesting ways
by substituting the bandgap $E_{g}$ with $2\left|\mu\right|$, where
$\mu$ is the chemical potential. The range of applicability of previous
theories can thus be handily tested by varying the doping level or
an applied gate voltage, and scenarios lying outside the typical operation
regime are readily accessible. In particular, for $\hbar\omega>2\left|\mu\right|$,
linear absorption of the fundamental beam occurs, which has not been
carefully considered in early studies of coherent optical injection
and control in carbon nanotubes \citep{Kral1999,Mele2000,Newson2008},
graphene \citep{Sun2010a,Rioux2011a,Sun2012b,Sun2012c}, and other
semiconducting materials \citep{Dupont1995,Atanasov1996,*Hache1997,Sheik-Bahae1999,Manykin2001,Marti2005a,Kral2007,Rodrigo2013}.

In the presence of coherent $\omega$ and $2\omega$ fields, linear
absorption at $\omega$ is not the only process with a net energy
absorption of $\hbar\omega$; at the next perturbative order the combined
absorption of a $2\omega$ photon and the emission of a $\omega$
photon must also be considered. This nondegenerate two-photon transition
process is an instance of stimulated electronic Raman scattering (ERS)
\citep{Wolff1966,Comas1986,Devereaux2007}, where light is scattered
inelastically and energy is deposited into the electronic state of
the system, and was recently studied in the nonlinear optical response
of graphene \citep{Cheng2014} and topological insulators \citep{Muniz2014}.
Since ERS at $\omega$ has the same initial and final state as 1PA
at $\omega$, quantum interference occurs between these two transition
pathways. Moreover, since the typical CC term between 1PA and 2PA
occurs at $2\omega$ rather than $\omega$, there are in effect two
distinct interference channels.

In this paper, we consider the carrier response to a two-color field
due to the ERS transition amplitude, first generally by presenting
a microscopic expression derived from Fermi's golden rule applicable
to condensed-matter systems, then specifically for the case of graphene.
For linearly-polarized light, it is shown that current injection in
graphene becomes anisotropic with respect to the angle between the
linear polarization axes of the light when the additional ERS contribution
is considered, and that this polarization sensitivity can be tuned
by adjusting the chemical potential of the sample. The current injection
resulting from the quantum interference between 1PA and ERS is five
times stronger with perpendicular polarization axes compared to parallel
polarization axes of the light. In contrast, the typical interference
term between 1PA and 2PA in graphene results in a current magnitude
independent of the polarization of the light \citep{Rioux2011a}.
The polarization sensitivity is adjustable, via the chemical potential,
between a completely isotropic current, to a strongly anisotropic
current response where the two contributions to current injection
are balanced to exactly cancel each other under perpendicular orientation
of the polarization axes.

We derive the microscopic form of the response tensors for carrier
and photocurrent injection via quantum interference of 1PA and ERS
in \prettyref{sec:New-contribution}. We consider the case of graphene,
explicitly calculate the response tensors, and present a complete
picture of two-color coherent current control via quantum interference
of one- and two-photon processes in this material in \prettyref{sec:Application-to-graphene}.
Finally, we conclude in \prettyref{sec:Conclusion}.

\section{Two-color interference at $\hbar\omega$ \label{sec:New-contribution}}

\begin{figure}
\capstart%

\includegraphics{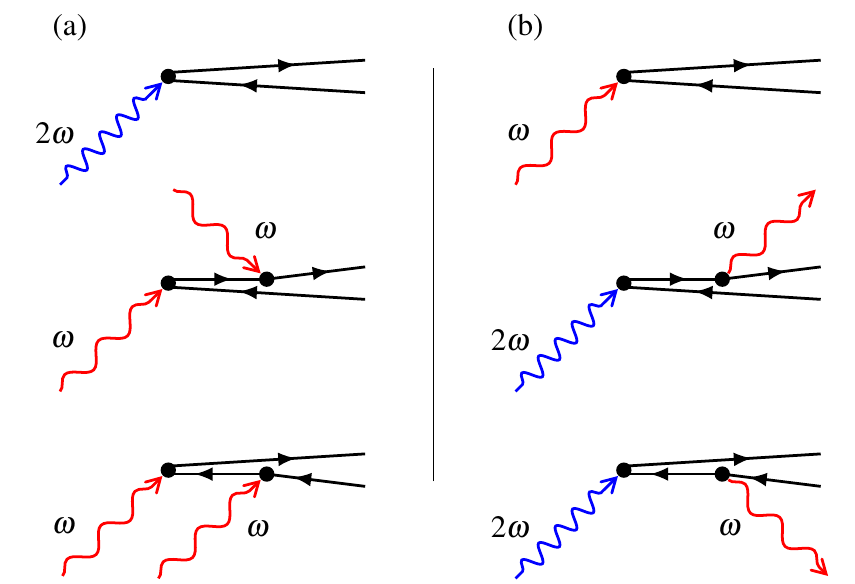}

\caption{Feynman diagrams of first- and second-order absorption processes contributing
to coherent control in the presence of a two-color, $\omega$ and
$2\omega$ field. The time axis points to the right, a plain arrow
pointing to the right (left) represents an electron (hole), and vertices
represent the dipole interaction. (a) Processes where the net energy
absorption amounts to $2\hbar\omega$: one-photon absorption (1PA)
at $2\omega$ and two-photon absorption (2PA) at $\omega$; the final
state is an electron-hole pair of energy $2\hbar\omega$. (b) Processes
where the net energy absorption amounts to $\hbar\omega$: 1PA at
$\omega$ and stimulated electronic Raman scattering (ERS) at $\omega$
($2\omega$ absorption and $\omega$ emission); the final state is
an electron-hole pair of energy $\hbar\omega$. The crossed processes,
where light is emitted before absorption, are also included in \prettyref{eq:Omega2}
but their diagrams omitted for brevity. Second-order processes where
the net energy absorption amounts to $3\hbar\omega$ or $4\hbar\omega$
are not included since they do not contribute to coherent control
unless third- and higher-order absorption processes are also included.
\label{fig:Feynman-diagrams} \label{subfiga:Feynman-diagrams-at-2omega}
\label{subfigb:Feynman-diagrams-at-omega}}
\end{figure}

In the presence of a two-color optical field with frequency components
$\omega$ and $2\omega$, the conventional injection current as originally
described by Atanasov \emph{et al.}~\citep{Atanasov1996} stems from
the cross-term of the following transition amplitudes connecting the
same initial and final states: $\Omega_{cv}^{(1)}(2\omega;\vec{k})$,
resulting from light at $2\omega$ to first order in perturbation,
and $\Omega_{cv}^{(2)}(\omega;\vec{k})$, resulting from light at
$\omega$ to second order in perturbation. Here $\Omega_{cv}^{(\ell)}(\omega;\vec{k})$
is the degenerate $\ell$-photon transition amplitude between valence
band $v$ and conduction band $c$ at wavevector $\vec{k}$ \citep{Rioux2012}.
The one- and two-photon absorption processes are illustrated by their
Feynman diagrams in \prettyref{subfiga:Feynman-diagrams-at-2omega}.

The injection term for the current density that results from the coherent
interference between the 1PA and 2PA transition amplitudes has the
form
\begin{equation}
\dot{J}^{a}=\eta_{I}^{abcd}(\omega)E^{b*}(\omega)E^{c*}(\omega)E^{d}(2\omega)+\cc,\label{eq:Jdot}
\end{equation}
where $\eta_{I}(\omega)$ is a fourth-rank tensor describing the current
response of the material, $\vec{E}(\omega)$ {[}$\vec{E}(2\omega)${]}
is the electric field component at frequency $\omega$ {[}$2\omega${]},
and repeated Roman superscripts are summed over Cartesian directions.
The tensor is given explicitly for a two-dimensional material by:
\begin{multline}
\eta_{I}^{abcd}(\omega)=\frac{\pi ie^{4}}{\hbar^{3}\omega^{3}}\sum_{c,v}\int\frac{\d^{2}k}{4\pi^{2}}[v_{cc}^{a}(\vec{k})-v_{vv}^{a}(\vec{k})]\\
\times w_{cv}^{bc*}(\vec{k})v_{cv}^{d}(\vec{k})\,\delta[\omega_{cv}(\vec{k})-2\omega]\label{eq:etaI}
\end{multline}
where $e=-\left|e\right|$ is the electron charge,
\begin{equation}
w_{cv}^{bc}(\vec{k})\equiv\sum_{m}\frac{v_{cm}^{b}(\vec{k})v_{mv}^{c}(\vec{k})+v_{cm}^{c}(\vec{k})v_{mv}^{b}(\vec{k})}{\omega_{mc}(\vec{k})+\omega_{mv}(\vec{k})},\label{eq:w}
\end{equation}
$\vec{v}_{mn}(\vec{k})\delta(\vec{k}-\vec{k}')=\bra{m\vec{k}}\vec{v}\ket{n\vec{k}'}$
indicate matrix elements of the velocity operator $\vec{v}$, $\omega_{cv}(\vec{k})\equiv\omega_{c}(\vec{k})-\omega_{v}(\vec{k})$,
and $\hbar\omega_{m}(\vec{k})$ are the band energies \citep{Atanasov1996,Rioux2011a}.

There is however another contribution to the current of the form of
\prettyref{eq:Jdot} that vanishes in a semiconductor when typical
frequencies are chosen to lie within the bandgap. This additional
term stems from interference between $\Omega_{cv}^{(1)}(\omega;\vec{k})$
describing one-photon absorption at $\omega$ and $\Omega_{cv}^{(2)}(2\omega,-\omega;\vec{k})$
describing the nondegenerate two-photon process of absorption at $2\omega$
and emission at $\omega$. These processes are illustrated by three
Feynman diagrams in \prettyref{subfigb:Feynman-diagrams-at-omega}.
The first diagram shows the one-photon absorption process; a $\omega$
photon is absorbed and an electron-hole pair is created with energy
$\hbar\omega$. This process has a $\vec{k}$-dependent transition
amplitude \citep{Rioux2012} given by
\begin{equation}
\Omega_{cv}^{(1)}(\omega;\vec{k})=\frac{ie}{\hbar\omega}v_{cv}^{a}(\vec{k})E^{a}(\omega)\label{eq:Omega1}
\end{equation}
and leads to the well-known linear absorption of $\omega$ light.
The second and third diagrams of \prettyref{subfigb:Feynman-diagrams-at-omega}
show the nondegenerate two-photon process; a $2\omega$ photon is
absorbed, a virtual electron-hole pair is created with energy $2\hbar\omega$,
and a $\omega$ photon is emitted by either the electron (second diagram)
or the hole (third diagram), reducing the energy of the electron-hole
pair to $\hbar\omega$. This is an instance of a stimulated electronic
Stokes-Raman process \citep{Wolff1966,Comas1986,Devereaux2007} known
to induce a two-color optical Kerr effect \citep{Sheik-Bahae1991,*Sheik-Bahae1992,*Sheik-Bahae2001}.
 We calculate the $\vec{k}$-dependent transition amplitude in the
single-particle approximation at the level of Fermi's golden rule,
and find
\begin{equation}
\Omega_{cv}^{(2)}(2\omega,-\omega;\vec{k})=\frac{e^{2}}{\hbar^{2}\omega^{2}}w_{cv}^{\prime ab}(\vec{k})E^{a*}(\omega)E^{b}(2\omega),\label{eq:Omega2}
\end{equation}
where
\begin{equation}
w_{cv}^{\prime ab}(\vec{k})\equiv\sum_{m}\left(\frac{v_{cm}^{a}(\vec{k})v_{mv}^{b}(\vec{k})}{\omega_{m}(\vec{k})-[\omega_{c}(\vec{k})+\omega]}+\frac{v_{cm}^{b}(\vec{k})v_{mv}^{a}(\vec{k})}{\omega_{m}(\vec{k})-[\omega_{v}(\vec{k})-\omega]}\right).\label{eq:wprime}
\end{equation}
The first term within the summation in \prettyref{eq:wprime} corresponds
to the second diagram of \prettyref{subfigb:Feynman-diagrams-at-omega},
the latter term corresponds to the third diagram.

The injection rate for the expectation value of any operator $\Theta$
is then calculated up to second perturbative order from $\dot{\left\langle \Theta\right\rangle }=i\hbar^{-1}\left\langle \left[\Hamiltonian,\Theta\right]\right\rangle $
using matrix elements of the Hamiltonian given by $\Hamiltonian_{mn}(\vec{k})=\hbar\omega_{m}(\vec{k})\delta_{mn}-\frac{e}{c}\vec{v}_{mn}(\vec{k})\cdot\vec{A}(t)$,
where $\vec{A}(t)$ is the vector potential, to obtain
\begin{align}
\dot{\left\langle \Theta\right\rangle } & =2\pi\sum_{c,v}\int\frac{\d^{2}k}{4\pi^{2}}\bra{c,v,\vec{k}}\Theta\ket{c,v,\vec{k}}\nonumber \\
 & \times\biggl(\left|\Omega_{cv}^{(1)}(\omega;\vec{k})+\Omega_{cv}^{(2)}(2\omega,-\omega;\vec{k})\right|^{2}\delta[\omega_{cv}(\vec{k})-\omega]\nonumber \\
 & \enskip\;+\,\left|\Omega_{cv}^{(1)}(2\omega;\vec{k})+\Omega_{cv}^{(2)}(\omega;\vec{k})\right|^{2}\delta[\omega_{cv}(\vec{k})-2\omega]\biggr).\label{eq:Thetadot}
\end{align}
The two summands within brackets describe electron-hole pairs of energy
$\hbar\omega$ and $2\hbar\omega$, respectively. The latter yields
the typical CC term described by \prettyref{eq:etaI} that has been
the subject of previous studies; we thus focus on the former. Taking
the occupation number operator as $\Theta$ yields the injection rate
for electron-hole pair creation at $\hbar\omega$,
\begin{align}
\dot{n}(\omega)\, & =\:\:\xi_{1}^{ab}(\omega)E^{a*}(\omega)E^{b}(\omega)\nonumber \\
 & +\Bigl[\xi_{I}^{\prime abc}(\omega)E^{a*}(\omega)E^{b*}(\omega)E^{c}(2\omega)+\cc\Bigr]\nonumber \\
 & +\hphantom{\Bigl[}\xi_{2}^{abcd}(2\omega,-\omega)E^{a*}(2\omega)E^{b*}(\omega)E^{c}(2\omega)E^{d}(\omega).\label{eq:ndot}
\end{align}
The first term is one-photon carrier injection due to 1PA, the third
term is nondegenerate, two-photon carrier injection (at $\omega$)
due to ERS, and the second term is their interference. The quantities
$\xi_{1}$, $\xi_{I}^{\prime}$ and $\xi_{2}$ are second-, third-
and fourth-rank tensors describing the material part of the response.
The 1PA carrier injection tensor is described elsewhere \citep{Atanasov1996}.
The response tensor describing the interference is given by
\begin{equation}
\xi_{I}^{\prime abc}(\omega)=\frac{2\pi ie^{3}}{\hbar^{3}\omega^{3}}\sum_{c,v}\int\frac{\d^{2}k}{4\pi^{2}}v_{cv}^{a*}(\vec{k})w_{cv}^{\prime bc}(\vec{k})\delta[\omega_{cv}(\vec{k})-\omega],\label{eq:xiIprime}
\end{equation}
and the ERS carrier injection tensor is given by
\begin{multline}
\xi_{2}^{abcd}(2\omega,-\omega)=\frac{2\pi e^{4}}{\hbar^{4}\omega^{4}}\sum_{c,v}\int\frac{\d^{2}k}{4\pi^{2}}w_{cv}^{\prime da*}(\vec{k})w_{cv}^{\prime bc}(\vec{k})\\
\times\delta[\omega_{cv}(\vec{k})-\omega].\label{eq:xi2}
\end{multline}
The interference term is analogous to coherent population control
of electron-hole pairs with net energy absorption of $2\hbar\omega$
in the conventional CC regime and is nonzero for noncentrosymmetric
materials \citep{Stevens2003b}.

Finally, taking the current operator as $\Theta$, the response tensor
describing the current injection resulting from interference between
ERS and 1PA at $\omega$ is found to be
\begin{multline}
\eta_{I}^{\prime abcd}(\omega)=\frac{\pi ie^{4}}{\hbar^{3}\omega^{3}}\sum_{c,v}\int\frac{\d^{2}k}{4\pi^{2}}[v_{cc}^{a}(\vec{k})-v_{vv}^{a}(\vec{k})]\\
\times\left[v_{cv}^{b*}(\vec{k})w_{cv}^{\prime cd}(\vec{k})+v_{cv}^{c*}(\vec{k})w_{cv}^{\prime bd}(\vec{k})\right]\,\delta[\omega_{cv}(\vec{k})-\omega].\label{eq:etaIprime}
\end{multline}
For a three-dimensional material, the integration over reciprocal
space in Eqs.~\eqref{eq:etaI}, \eqref{eq:Thetadot}, \eqref{eq:xiIprime},
\eqref{eq:xi2} and \eqref{eq:etaIprime} should read $\int\d^{3}k/8\pi^{3}$.

\section{Anisotropic photoinduced current injection in graphene \label{sec:Application-to-graphene}}

\begin{figure}
\capstart%

\includegraphics{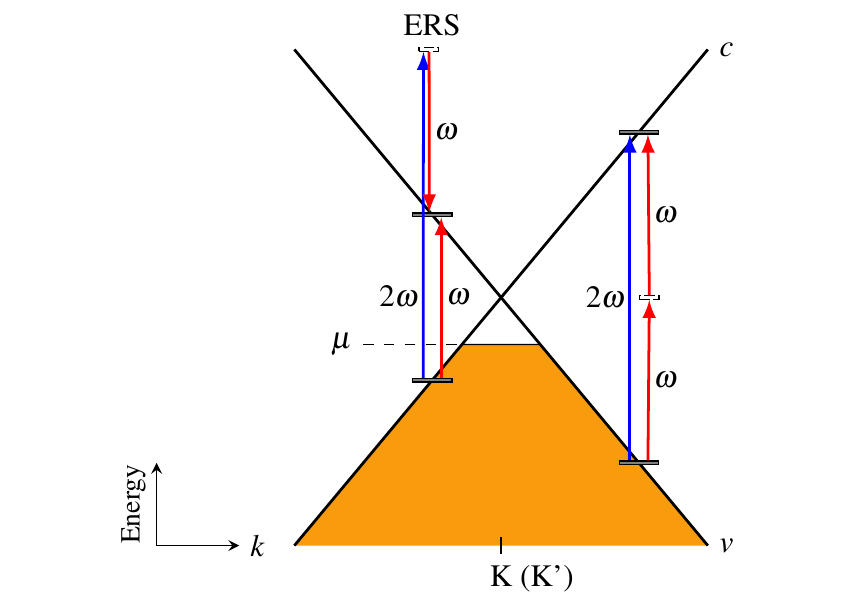}

\caption{The conventional excitation scheme (right) employs interference between
two-photon absorption at $\omega$ (red arrows) and one-photon absorption
at $2\omega$ (blue arrows). The additional contribution at $\hbar\omega$
(left) occurs when one-photon absorption at $\omega$ interferes with
stimulated electronic Raman scattering (ERS, $2\omega$ absorption
and $\omega$ emission). Depending on the chemical potential $\mu$,
the ERS contribution is allowed ($2\left|\mu\right|<\omega$) or blocked
($2\left|\mu\right|>\omega,$ not shown). \label{fig:graphene-band-structure-and-excitation-scheme}}
\end{figure}

We consider graphene to illustrate the two-color interference effect
at $\hbar\omega$ in experimentally accessible conditions. The part
of reciprocal space relevant for optical response are the two valleys
near K and K' at the vertices of the hexagonal Brillouin zone, where
the electrons follow a Dirac-like spectrum {[}\prettyref{fig:graphene-band-structure-and-excitation-scheme}{]}.
The gapless band dispersion has the advantage that a bandgap energy
is not a limiting factor for the occurrence of absorption at $\hbar\omega$.
Instead, the chemical potential $\mu$ plays the role of the limiting
factor. Photocurrent generation via CC includes the contribution from
both interference channels presented in the previous section and illustrated
in \prettyref{fig:graphene-band-structure-and-excitation-scheme}
for graphene. Experimentally, electrically tuning the chemical potential
allows one to enter and exit the regime where absorption at $\hbar\omega$
occurs. It is thus possible to identify the ERS contribution.

The diagonalized model Hamiltonian and corresponding velocity operator
near the K point of graphene take the form \citep*{Rioux2011a}
\begin{align}
\Hamiltonian_{0} & \rightarrow\hbar v_{F}\left(\begin{array}{cc}
k & 0\\
0 & -k
\end{array}\right), & \vec{v} & \rightarrow v_{F}\left(\begin{array}{cc}
\hat{k} & i\hat{\phi}\\
-i\hat{\phi} & -\hat{k}
\end{array}\right),
\end{align}
where $v_{F}$ is the Fermi velocity, $\vec{k}=k\hat{k}$ is the in-plane
crystal momentum relative to the Dirac point, and $\hat{\phi}=\hat{z}\times\hat{k}$
is an in-plane unit vector perpendicular to $\vec{k}$. The treatment
for the K' valley is analogous and its presentation neglected for
the remainder of the paper as it simply introduces a degeneracy factor
of two, in addition to the spin degeneracy factor of two.

The generation of a photocurrent via CC is best illustrated by the
unbalanced carrier distributions in reciprocal space due to the interference
effect being constructive or destructive at different wavevectors
\citep{Sheik-Bahae1999}. The carrier distributions due to 2PA and
1PA at $2\omega$ have been previously calculated \citep{Rioux2011a}.
Here we present the result of the ERS contribution and its interference
with 1PA at $\omega$. The ERS has previously been calculated in the
context of conventional Raman spectroscopy \citep{LuWang2008}.

\begin{figure}
\capstart%

\includegraphics{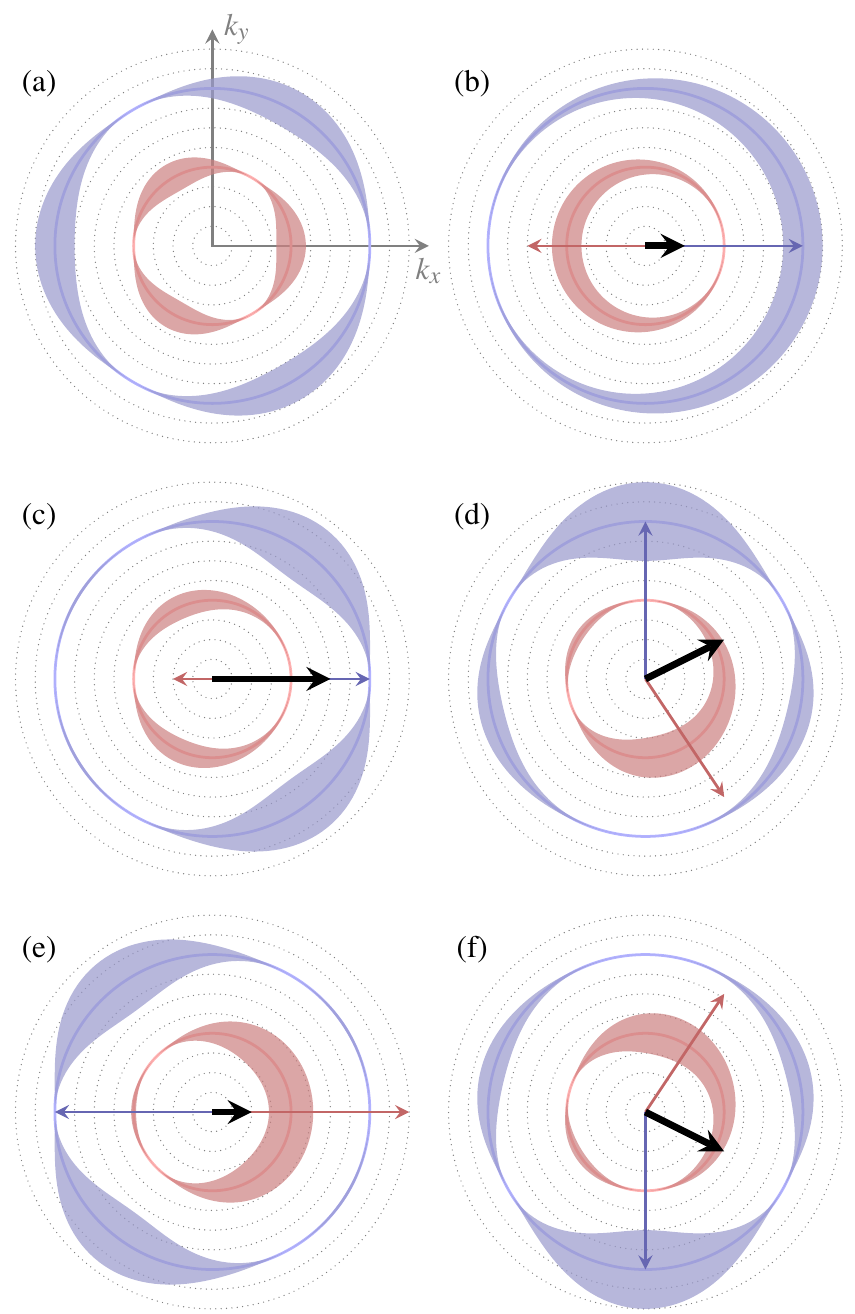}

\caption{Reciprocal space distribution of injected carriers in intrinsic graphene
under two-color, $\omega$ and $2\omega$ field. Absorption processes
up to second order in the field are considered, with net energy absorption
of $\hbar\omega$ (red, inner distribution) and $2\hbar\omega$ (blue,
outer distribution), respectively due to ERS-1PA interference and
1PA-2PA interference. The carrier injection rates {[}\prettyref{eq:ndot@2omega}
and \prettyref{eq:ndot@omega}{]} are indicated by the thickness of
the distributions, the current resulting from each distribution is
indicated by a correspondingly colored arrow, and the overall current
injection is indicated by a thick, black arrow. Photoresponse for
circularly-polarized $\omega$ and $2\omega$ beams: (a) opposite
polarization, yielding no net current injection, and (b) co-circular
polarization, yielding a net current injection. Photoresponse for
linearly-polarized $\omega$ and $2\omega$ beams: (c) $\theta=0$,
\emph{i.e.}~parallel polarization axes, (d) $\theta=\pi/4$, (e)
$\theta=\pi/2$, \emph{i.e.}~perpendicular polarization axes, and
(f) $\theta=3\pi/4$; all yielding current injection. \label{fig:distributions}
\label{subfige:distributions}}
\end{figure}

The distribution of photoexcited carriers in reciprocal space due
to the two-color field, $\dot{n}=\dot{n}(2\omega;\vec{k})+\dot{n}(\omega;\vec{k})$,
is shown in \prettyref{fig:distributions}, taking both the conventional
and ERS contributions to be equally allowed, e.g.~for intrinsic graphene.
The distribution due to conventional CC is
\begin{equation}
\dot{n}(2\omega;\vec{k})=\left|\Omega_{cv}^{(1)}(2\omega;\vec{k})+\Omega_{cv}^{(2)}(\omega;\vec{k})\right|^{2}\label{eq:ndot@2omega}
\end{equation}
and occurs on the (outer) circle of radius $k=\omega/v_{F}$; the
distribution due to ERS-induced CC is
\begin{equation}
\dot{n}(\omega;\vec{k})=\left|\Omega_{cv}^{(1)}(\omega;\vec{k})+\Omega_{cv}^{(2)}(2\omega,-\omega;\vec{k})\right|^{2}\label{eq:ndot@omega}
\end{equation}
and occurs on the (inner) circle of radius $k=\omega/2v_{F}$.

For circular polarization {[}\prettyref{fig:distributions}(a,b){]}
the distributions depend on the angle $\phi_{k}=\tan^{-1}(k_{y}/k_{x})$
and the CC parameter $\Delta\varphi\equiv2\varphi_{\omega}-\varphi_{2\omega}$
relating the phases of the two frequency components. For opposite-circularly
polarized $\omega$ and $2\omega$ beams, $\hat{\vec{e}}_{2\omega}=-\hat{\vec{e}}_{\omega}=\sigma^{\pm}$,
the distributions for the carrier injection with net energy absorption
of $2\hbar\omega$ and $\hbar\omega$ vary as $\dot{n}(2\omega;\vec{k})\propto1-\sin(\Delta\varphi\mp3\phi_{k})$
and $\dot{n}(\omega;\vec{k})\propto\tfrac{3}{4}\big(1+\sin(\Delta\varphi\pm3\phi_{k})\big)$,
respectively, neither of which yield a current. For co-circularly
polarized beams, $\hat{\vec{e}}_{2\omega}=\hat{\vec{e}}_{\omega}=\sigma^{\pm}$,
the distributions of $\dot{n}(2\omega;\vec{k})\propto1+\sin(\Delta\varphi\pm\phi_{k})$
and $\dot{n}(\omega;\vec{k})\propto\tfrac{3}{4}\big(1-\sin(\Delta\varphi\mp\phi_{k})\big)$
yield opposite current contributions, although a net current injection
remains. The distributions are plotted for $\Delta\varphi=\pi/2$.

For linearly-polarized beams {[}\prettyref{fig:distributions}(c-f){]},
without loss of generality the polarization axis $\hat{\vec{e}}_{2\omega}$
is chosen along $\hat{x}$ and the $\vec{k}$-dependent carrier distributions
follow
\begin{align}
\dot{n}(2\omega;\vec{k}) & \propto\left|\sin(\phi_{k})+ie^{-i\Delta\varphi}\sin(2\phi_{k}-2\theta)\right|^{2},\label{eq:ndotlinear@2omega}\\
\dot{n}(\omega;\vec{k}) & \propto\left|\sin(\phi_{k}-\theta)-\tfrac{i}{4}e^{-i\Delta\varphi}\left[\sin(2\phi_{k}-\theta)+3\sin(\theta)\right]\right|^{2},\label{eq:ndotlinear@omega}
\end{align}
where $\theta$ is the angle between the polarization axes $\hat{\vec{e}}_{\omega}$
and $\hat{\vec{e}}_{2\omega}$. The first term in each of Eqs.~\eqref{eq:ndotlinear@2omega}
and~\eqref{eq:ndotlinear@omega} is due to 1PA, and the second term
is due to 2PA {[}\prettyref{eq:ndotlinear@2omega}{]} or stimulated
ERS {[}\prettyref{eq:ndotlinear@omega}{]}. The distributions are
shown for $\Delta\varphi=\pi/2$ and the polarization axis $\hat{\vec{e}}_{\omega}$
forming angles of $\theta=0$, $45^{\circ}$, $90^{\circ}$, and $135^{\circ}$
with respect to $\hat{\vec{e}}_{2\omega}$. For parallel and perpendicular
polarization, the current due to the ERS contribution opposes the
current due to the conventional injection process; in the latter case,
the ERS contribution is strong enough to cause the direction of the
net current to flip. For $45^{\circ}$ and $135^{\circ}$, the ERS
contribution adds a current component along the $\hat{\vec{e}}_{2\omega}$
direction, while its component along $\hat{\vec{e}}_{2\omega}^{\perp}$
opposes the conventional injection process, although not sufficiently
to flip the net current along this direction. The magnitude of the
photocurrent is anisotropic in $\theta$ due to the ERS contribution.

The symmetry of graphene yields $\eta_{I}^{xxxx}$, $\eta_{I}^{xyyx}$,
and $\eta_{I}^{xyxy}=\eta_{I}^{xxyy}$ as independent components of
the current injection tensor; in addition, the relation $2\eta_{I}^{xyxy}=\eta_{I}^{xxxx}-\eta_{I}^{xyyx}$
holds since an isotropic band model is considered. The disparity parameter
$\disparity=\eta_{I}^{xyyx}/\eta_{I}^{xxxx}$ characterizes the sensitivity
of the current injection to the angle $\theta$ between polarization
axes for linearly-polarized $\omega$ and $2\omega$ beams; different
values of $d$ and $\theta$ lead to injected currents with different
magnitude and direction \citep*{Rioux2011a}.  The additional current
injection term presented in this paper yields the following implications
for graphene. In the conventional regime {[}\prettyref{eq:etaI}{]},
the nonzero current injection tensor components were given by
\begin{equation}
\eta_{I}^{xxxx}=\eta_{I}^{xyxy}=\eta_{I}^{xxyy}=-\eta_{I}^{xyyx}=i\bar{\eta}_{I}(\omega),
\end{equation}
where
\begin{equation}
\bar{\eta}_{I}(\omega)\equiv g_{s}g_{v}e^{4}v_{F}^{2}\left(2\hbar\omega\right)^{-3},
\end{equation}
with $g_{s}=g_{v}=2$ describing spin and valley degeneracy, respectively
\citep{Rioux2011a}. As a result of this contribution, a current component
is injected with effective parameter $\disparity_{\text{eff}}=-1$
and the magnitude of this current is $\theta$-independent. The ERS
contribution, in contrast, is strongly $\theta$-dependent: taken
on its own it contributes an effective parameter $\disparity_{\text{eff}}=-5$,
meaning that it yields a current whose magnitude is five times stronger
when the beam polarization axes are perpendicular ($\theta=\pi/2$)
compared to them being parallel ($\theta=0$). The nonzero tensor
components for this contribution {[}\prettyref{eq:etaIprime}{]} are
given by
\begin{equation}
\eta_{I}^{\prime xxxx}=\tfrac{1}{3}\eta_{I}^{\prime xyxy}=\tfrac{1}{3}\eta_{I}^{\prime xxyy}=-\tfrac{1}{5}\eta_{I}^{\prime xyyx}=-\tfrac{1}{4}i\bar{\eta}_{I}(\omega).
\end{equation}
This contribution doesn't occur if the transition at $\omega$ is
blocked due to the lower level being depleted or the upper level occupied.
For intrinsic graphene the two discussed interference channels are
equally allowed, the resulting current injection tensor includes both
contributions, its components are given by
\begin{equation}
\eta_{I}^{xxxx}=3\eta_{I}^{xyxy}=3\eta_{I}^{xxyy}=3\eta_{I}^{xyyx}=\tfrac{3}{4}i\bar{\eta}_{I}(\omega),
\end{equation}
and the corresponding parallel-perpendicular polarization disparity
parameter is $\disparity=1/3.$ This value is closer to a typical
semiconductor such as GaAs, which has $\left|\disparity\right|\approx0.2$
for a fundamental photon energy of about 1\,eV (below the onset of
1PA at $\omega$) \citep{Rioux2012}. For intrinsic graphene the parameter
$\disparity$ is independent of the light frequency, as it would be
were the ERS contribution not included \citep{Rioux2011a}.

\begin{figure}
\capstart%

\includegraphics{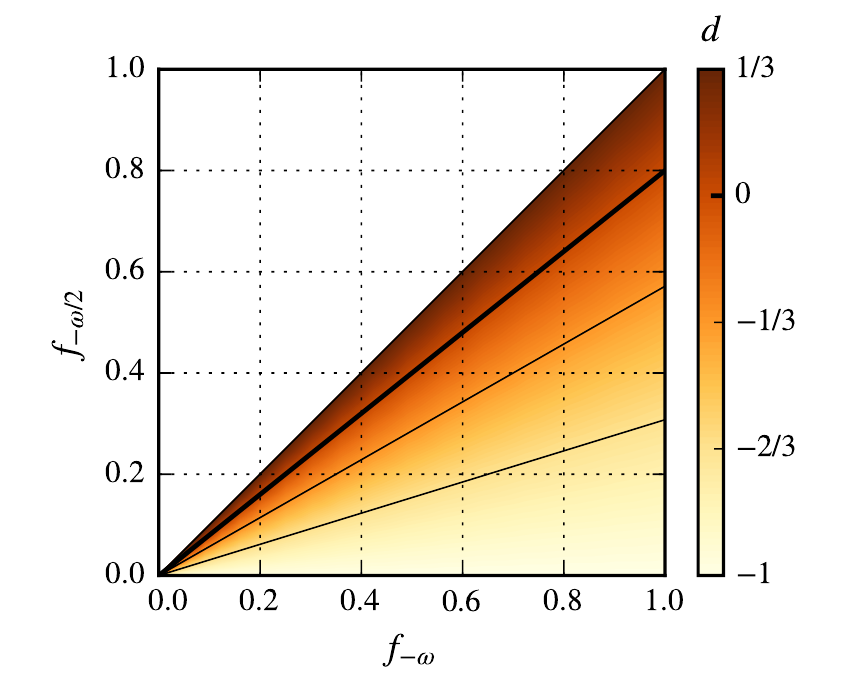}

\caption{Contour plot of the parallel-perpendicular polarization disparity
parameter $\disparity=\eta_{I}^{xyyx}/\eta_{I}^{xxxx}$ in $p$-doped
graphene as a function of the electron population $f_{-\omega}$ and
$f_{-\omega/2}$ in the valence band ($f_{-\omega/2}\le f_{-\omega}$).
With $f_{-\omega/2}=0$, stimulated ERS and 1PA at $\omega$ is blocked
and interference of 2PA and 1PA at $2\omega$ yields $d=-1$ characterizing
an injection current that is isotropic with respect to $\theta=\cos^{-1}(\hat{\vec{e}}_{\omega}\cdot\hat{\vec{e}}_{2\omega})$.
With $f_{-\omega/2}\ne0$, the additional ERS contribution yields
$d>-1$ characterizing an anisotropic current. In particular, for
$d=0$ (thick contour line) the injection current vanishes for perpendicular
polarization axes, $\theta=\pi/2$. \label{fig:contour-plot}}
\end{figure}

Let us now consider the effect of a finite temperature and nonzero
chemical potential on the current injection. The deviation from the
semiconductor vacuum state results in the Pauli blocking of optical
transitions, as observed in linear absorption \citep{Mak2008}. Since
the two contributions to current injection, \prettyref{eq:etaI} and
\prettyref{eq:etaIprime}, occur at different excitation energies,
it is possible to partly block the lower-energy transition while retaining
the higher-energy one. This allows one to tune the polarization sensitivity
of the resulting current. For a $p$-doped sample, if $f_{\omega}$
is the population of electrons as a function of the energy level $\hbar\omega$
relative to the Dirac point, then the disparity parameter is
\begin{equation}
\disparity=-\frac{4f_{-\omega}-5f_{-\omega/2}}{4f_{-\omega}-f_{-\omega/2}}.
\end{equation}
This expression is plotted in \prettyref{fig:contour-plot}, and shows
that the allowed values are in the range $-1\le\disparity\le1/3$.
Taking $f_{-\omega/2}=0$ yields $\disparity=-1$ and corresponds
to the previous result with complete blocking of the linear absorption
of the fundamental due to the lower level being depleted \citep{Rioux2011a},
while the ratio of $f_{-\omega/2}/f_{-\omega}=1$ yields $d=1/3$
and corresponds to the situation presented in \prettyref{fig:distributions}.
For $d>0$, the injected current in the cross-polarized configuration
flips direction due to the ERS contribution. A value of $d=0$ (thick
contour line in \prettyref{fig:contour-plot}) corresponds to the
complete suppression of the current injection for perpendicular polarization
axes; this is achieved by adjusting the population to the ratio of
$f_{-\omega/2}/f_{-\omega}=0.8$. Thus, a range of polarization-sensitive
and polarization-insensitive configurations are available.

\begin{figure}
\capstart%

\includegraphics{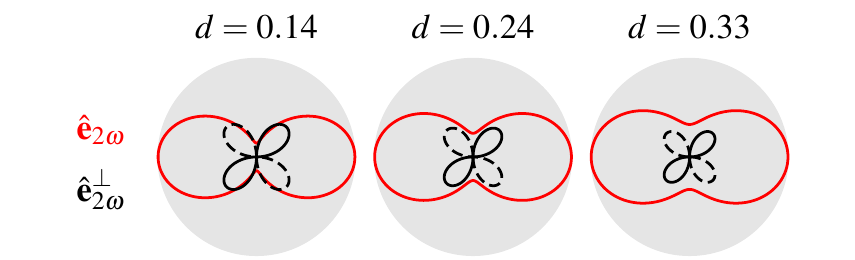}

\caption{Angular plots of the current injection strength as a function of the
angle $\theta$ between the linear polarization axes $\hat{\vec{e}}_{\omega}$
and $\hat{\vec{e}}_{2\omega}$, for the ratio $d=\eta_{I}^{xyyx}/\eta_{I}^{xxxx}$
taking values of $0.14$, $0.24$, and $0.33$. The current components
along $\hat{\vec{e}}_{2\omega}$ (plain red line) and $\hat{\vec{e}}_{2\omega}^{\perp}$
(plain/dashed black line) are plotted; a dashed line indicates a negative
current along that axis. \label{fig:angular-plots}}
\end{figure}

The variation of the current injection characteristics according to
temperature and nonzero chemical potential affects our understanding
of previous experiments in multilayer epitaxial graphene \citep{Sun2010a,Sun2012b,Sun2012c},
as the contribution from a single layer depends on its doping level.
The doping level in multilayer epitaxial graphene decreases exponentially
from 365\,meV for the layer closest to the substrate to zero doping
for the top layers, with a charge screening length of one layer \citep{Sun2010a}.
The total current injection across all layers is made up of an isotropic
contribution from the few heavily-doped layers near the substrate
($d=-1)$ and an anisotropic contribution from the remaining mostly
undoped layers ($d=1/3$). In the original experiment the incoming
beams were cross-linearly polarized \citep{Sun2010a}; taking the
ERS contribution into account yields counter-propagating currents
in the heavily-doped and undoped layers, as illustrated in \prettyref{subfige:distributions}.
A subsequent experiment resolved the angular dependence on the polarization
axes and found that it did not correspond to the simple model of a
single graphene layer \citep{Sun2012b}. In both experiments the current
injection was detected via the emitted THz radiation with the signal
collected behind the sample. Light attenuation is stronger for the
THz signal than for the frequencies of the incoming beams. Thus, although
there are only a few heavily-doped layers, they could contribute relatively
strongly to the detected THz signal due to their proximity with the
detector. It is realistic to expect the signal to have a value of
$d$ in the range $0.14$ to $0.24$; the resulting polarization dependence
is shown in \prettyref{fig:angular-plots}, and is in good agreement
with the polarization-resolved experimental data \citep{Sun2012b}.
Thus the inclusion of the Raman term identified here could lead to
an understanding of the experimental results even without the assumption
of interlayer coupling.

\section{Conclusions \label{sec:Conclusion}}

We have presented the complete current response of the gapless semiconductor
graphene due to coherent control of light at frequencies $\omega$
and $2\omega$. By considering the additional interference effect
between stimulated ERS and 1PA of the fundamental, occurring with
a net energy absorption of $\hbar\omega$, we have shown that graphene
presents a photoresponse sensitive to the relative orientation of
the linear polarization axes of the light. This polarization sensitivity,
characterized by a tunable disparity parameter $-1\le\disparity\le1/3$,
is in stark contrast with the polarization-isotropic photocurrent
contribution due to 1PA and 2PA interference alone \citep{Rioux2011a}.
The value of $d=1/3$ for intrinsic graphene is closer to typical
values in GaAs \citep{Rioux2012}, and with a value of $d=0$ it is
possible to completely suppress the current injection for perpendicular
polarization axes. Thus, with its Fermi level adjustable through doping
or electrical gating, graphene offers tunable, polarization sensitive
applications.

Although the excitation mechanism presented here differs from the
excitation mechanisms of previous studies, it is expected that the
subsequent dynamics follows a similar course. The injected photocurrent
is limited by momentum relaxation and is expected to decay following
excitation. For a pulsed laser source, it has been reported that hot
carriers reach an isotropic distribution 150\,fs to 250\,fs after
excitation, with intraband carrier-phonon scattering being the main
momentum relaxation mechanism \citep{Dawlaty2008a,Breusing2011,Winnerl2011,Sun2012c,Brida2013,Mittendorff2014}.
On shorter timescales, collinear carrier-carrier scattering preserves
the anisotropy \citep{Sun2012c,Brida2013,Mittendorff2014}, allowing
for similar photocurrents to be detected \citep{Sun2010a,Sun2012b,Sun2012c}.

While the additional interference term between stimulated ERS and
1PA is correctly included in a recent treatment of third-order nonlinear
optical conductivities of doped graphene \citep{Cheng2014}, it had
not been carefully considered in previous studies of coherent optical
injection and control in graphene and carbon nanotubes \citep{Kral1999,Mele2000,Newson2008,Sun2010a,Rioux2011a,Sun2012b,Sun2012c}.
However, we have shown that this term modifies the CC photocurrent
response significantly, raising the question whether previous experimental
results need to be reinterpreted \citep{Newson2008,Sun2010a,Sun2012b,Sun2012c}.
In experiments with fixed polarization the ERS contribution does not
play a crucial role, specifically the conclusions of Newson \emph{et
al.}\ regarding CC using colinearly-polarized beams in carbon nanotubes
\citep{Newson2008}, and those of Sun \emph{et al.}\ using cross-polarized
beams in multilayer epitaxial graphene \citep{Sun2010a,Sun2012c},
are not altered. Considering the ERS contribution and depending on
the level of doping, cross-polarized beams could lead to counter-propagating
currents in the undoped and heavily-doped layers of epitaxial graphene,
an effect that is not resolvable in the employed THz detection technique.
Such a different response from the undoped and heavily-doped layers
offers an alternative explanation for the polarization-sensitive data
of Sun \emph{et al}.~{[}\citealp{Sun2012b}{]}, which the authors
attributed to interlayer coupling.

Besides gapless semiconductors such as graphene, a careful treatment
of the interference between stimulated electronic Raman scattering
and linear absorption is important for the complete description of
two-color coherent control in metals, topological insulators, and
in semiconductors when $\hbar\omega$ exceeds the bandgap. Calculations
like the one we presented will play an important role in the analysis
of experiments involving those systems.

\appendix

\begin{acknowledgments}
This work was supported by the Deutsche Forschungsgemeinschaft (DFG)
under project number SFB 767 and the Konstanz Center for Applied Photonics
(CAP). J.\,E.\,S. acknowledges funding from the National Sciences
and Engineering Research Council (NSERC) of Canada. The authors acknowledge
useful discussions with Jin Luo Cheng, Dong Sun and Ted Norris.
\end{acknowledgments}

\def\firstpage#1--#2\relax{#1}\def\pages#1{\firstpage
  #1\relax}

\end{document}